\documentclass{kapproc} 

\usepackage{procps} 

\usepackage[dvips]{graphicx}

\upperandlowercase

\kluwerbib 

\begin{document}

\articletitle{Review on the Observed and Physical Properties of Core Collapse Supernovae}


\author{Mario Hamuy}
\affil{Carnegie Observatories}
\email{mhamuy@ociw.edu}

\begin{abstract}
Core collapse supernovae prove to comprise the most common general class of
exploding star in the Universe and they come in a great variety of flavors.
The wide range of luminosities, expansion velocities, and chemical abundances
displayed by these objects is evidence for large variations in explosion energy
and in the properties of their progenitors. This paper summarizes observed
and physical properties of all types of core collapse supernovae.
Despite the great diversity displayed by these objects, several regularities
emerge which suggest that 1) there is a continuum in the properties of these
objects, 2) the mass of the envelope is one of the driving parameters of the
explosion, or it is correlated with some other property of the core, with
the latter determining the outcome of the explosion, and 3) the physics of
the core and explosion mechanism of all core collapse supernovae are not be
fundamentally different, regardless of the external appearance of the supernova.
Far above in energy scale and $^{56}$Ni production lies SN~1998bw, the only supernova
firmly associated with a gamma-ray burst.
\end{abstract}


\section*{Introduction}

Supernovae (SNe) owe their name to astronomers Baade and Zwicky who, in the
1930's, realized that these objects were much more luminous and rarer than
common novae (\cite{Baade38,Zwicky38}). Their high luminosities 
(comparable to that of their host galaxies) and broad spectral lines 
led them to conclude that SNe were very energetic explosions. They went a step
further and hypothesized that a SN resulted from the ``transformation of
an ordinary star into a collapsed neutron star'', a remarkable idea for
its time which lies at the heart of modern models for SNe that
result from the gravitational collapse of the cores of massive stars.

``Core collapse SNe'' (CCSNe, hereafter) prove to comprise the most
common general class of exploding star in the Universe (\cite{Cappellaro99}), each releasing
$\sim$10$^{51}$ ergs of mechanical energy and enriching the interstellar medium
with several solar masses of new chemical elements.  Their astrophysical
importance is no longer limited to the central role they play in the chemical
evolution of the Universe and in the shaping of the galaxies themselves,
but now extends to the possibility that a fraction of them might be the source
of the enigmatic gamma-ray bursts (GRBs).

CCSNe come in a great variety of flavors. In this paper, I summarize
the photometric and spectroscopic properties of all types of CCSNe and their
physical parameters. Despite the great diversity displayed by these objects several
regularities emerge which provide valuable clues and a better insight on their
explosion mechanism, a matter that still remains quite controversial
(see \cite{Janka03}, \cite{Cardall03}, and \cite{Burrows00} for recent reviews).

\section{Supernova Classification} 
\label{classification}

Observers classify SNe according to the presence or absence of certain
elements in their atmospheres based on spectroscopic observations.
By the time Baade and Zwicky introduced the SN class, spectra of
these objects had already been obtained. \cite{minkowski41} published the first paper
on this subject where he introduced two main SN spectroscopic types:
the Type II class comprises SNe with prominent hydrogen lines, whereas
the Type I class is defined by the absence of hydrogen in their spectra.
This classification scheme has evolved as more spectra have become
available (see \cite{Filippenko97} for a detailed review).
Five distinct SN types can be distinguished from 
spectra obtained near maximum light (Fig. \ref{SC.fig}):

\begin{figure}[ht]
\centerline{\includegraphics[width=3.3in,height=3.3in]{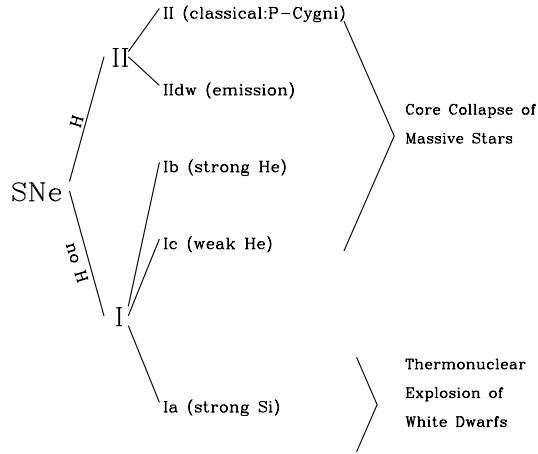}}
\caption{Supernova Classification.}
\label{SC.fig}
\end{figure}

\noindent $\bullet$ classical Type II: These objects have prominent Balmer lines
exhibiting P-Cygni profiles. 

\noindent $\bullet$ Type IIdw: The members of this class have strong
hydrogen lines in emission. They can be distinguished from the classical Type II SNe
by the lack of absorption in their Balmer lines. \cite{chugai97a} introduced
this designation to reflect the fact that these SNe undergo significant
interaction with a ``dense wind'' produced by the SN progenitor prior to explosion.

\noindent $\bullet$ Type Ia: They are characterized by a strong absorption
attributed to Si II $\lambda$$\lambda$ 6347,6371.

\noindent $\bullet$ Type Ib: These objects are distinguished by spectra with
no evident Balmer lines, weak or absent Si II $\lambda$$\lambda$ 6347,6371,
and strong He I $\lambda$$\lambda$ 4471, 5876, 6678, and 7065 lines.
\cite{bertola64} reported the first observations of this class of SNe
but the ``Ib'' designation was introduced later by \cite{elias85}.

\noindent $\bullet$ Type Ic: The members of this class are characterized
by weak or absent hydrogen and helium lines, and no evident Si II $\lambda$$\lambda$ 6347,6371.
They show Ca II H\&K in absorption, the Ca II near-IR triplet with
a P-Cygni profile, and O I $\lambda$ 7774 in absorption. The ``'Ic''
class was introduced by \cite{wheeler86}.

During the past years a new class of SN seems to be emerging, which
is characterized by a smooth and featureless spectrum at early epochs.
The current interpretation is that these objects have the usual lines
observed in SNe~Ic but with an extreme Doppler broadening
caused by unusually high expansion velocities. There are
three members of this class (SN~1997ef, SN~1998bw, and SN~2002ap),
which are often called ``hypernovae'' or SNe~Id. 
One of them (SN~1998bw) proved to be a remarkable
event because it was found at the same temporal and spatial location
as GRB980425 (\cite{Galama98}). In the rest of this paper I will
refer to these objects as Type Ic hypernovae to reflect
the observational fact that their expansion velocities are unusually high.

SNe~II, Ib, and Ic occur near star forming regions and
have never been observed in elliptical galaxies, which leads
to the idea that their progenitors are massive stars born 
with more than $\sim$8 $M_\odot$ that undergo core collapse,
leaving a neutron star or black hole as a remnant and launching
an explosion of their envelopes.
Type Ia SNe, on the other hand, are observed in all types
of galaxies. Given their lack of hydrogen, it is thought that
they arise from white dwarfs that explode as they approach the Chandrasekhar mass
($\sim$ 1.4 $M_\odot$) after a period of mass accretion from a binary companion,
leaving no compact remnant behind them. 

Theorists give less importance to the external appearance of SNe (spectra) but
to their hearts (the origin of the explosion), and distinguish two fundamentally different SN types
regardless of their spectroscopic appearance: core collapse and thermonuclear SNe.
In this sense SNe~Ib and SNe~Ic are thought to be physically much closer to SNe~II
than to SNe~Ia, even though SNe Ia, Ib, and Ic all share the same prefix
(owing to the lack of hydrogen in their atmospheres).

\section{The Properties of Core Collapse Supernovae}

Stars born with $\sim$8-10 $M_\odot$ can reach temperatures
and densities sufficiently high to produce O/Ne/Mg cores.
More massive stars can proceed even further and end up with
Fe nuclei. When nuclear burning ceases in these stars the core
becomes unstable and gravitational collapse follows, leading to
the formation of a neutron star or black hole. It is thought that
this is the place where a supernova is launched, but the mechanism
by which the core implosion triggers the explosion of the stellar
mantle still remains a difficult theoretical problem (see \cite{Janka03}, for example).
The standard paradigm is that most of the gravitational energy released
during the collapse is carried away by neutrinos ($\sim$10$^{53}$ ergs) and that a small
fraction ($\sim$1\%) of this energy is deposited in the
bottom of the star's envelope. This produces a shock wave that
propagates through the interior of the star and emerges on
the surface a few hours later. This is a generic model
for CCSNe, whose optical luminosities comprise only
a small fraction of the energy released by the gravitational collapse, and
whose optical spectra vary solely by the ability of their
progenitors to retain their H-rich and He-rich envelopes
prior to explosion and/or by the density of the medium
in which they explode. In what follows I proceed to discuss the
properties of all spectroscopic types of CCSNe.

\subsection{ Classical Type II Supernovae }

\begin{figure}[ht]
\centerline{\includegraphics[width=3.3in,height=3.3in]{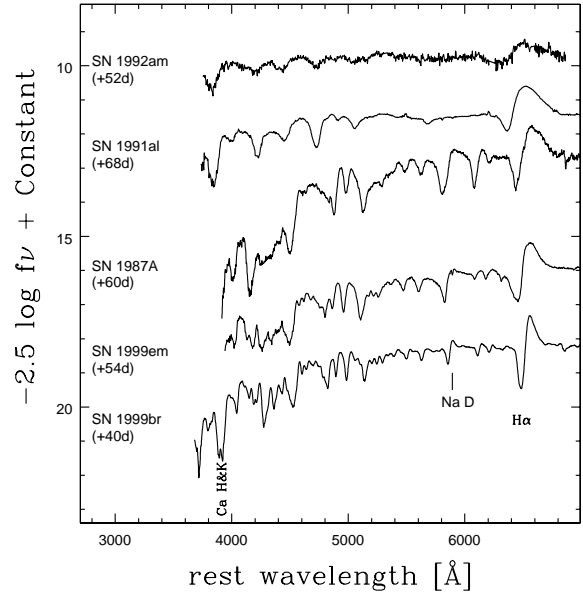}}
\caption{Optical spectra of classical Type II SNe obtained $\sim$60 days after explosion,
sorted by the width of their spectral lines. The spectrum of SN~1987A
is from \cite{phillips88}, that of SN~1999em is from \cite{leonard02a},
and the rest are from \cite{hamuy01a}.}
\label {spec1.fig}
\end{figure}

These SNe are thought to have massive progenitors with extended H-rich
envelopes that undergo little interaction with the circumstellar medium (CSM).
They are characterized by optical spectra dominated by Balmer lines exhibiting
P-Cygni profiles. Many of these SNe have been aimed with radio telescopes, yet
only a handful of nearby events such as SN~1987A and SN~1999em have been
detected at these frequencies (\cite{Weiler02}). The low radio luminosity of
these objects implies that they explode in low-density environments.
Based on the LOTOSS (\cite{Li00}) discoveries reported to the IAU Circulars in 2001,
I estimate that classical SNe~II constitute $\sim$45\% of all CCSNe.

Figure \ref{spec1.fig} displays optical spectra of a sample of classical SNe~II
taken $\sim$60 days after explosion. The most prominent feature of these
objects is H$\alpha$ exhibiting a P-Cygni profile, which is characteristic
of an expanding atmosphere (\cite{Kirshner74,Jeffery90}). Besides the Balmer
series these spectra show strong Ca H\&K, Na I D, and Fe lines.
Clearly this sample of SNe reveals a great degree of individuality.
In particular it is possible to observe a wide range in the line widths, from
the ``narrow-line'' SN~1999br to the ``broad-line'' SN~1992am, which suggests
a significant range in expansion velocities.

\begin{figure}[ht]
\centerline{\includegraphics[width=3.3in,height=3.3in]{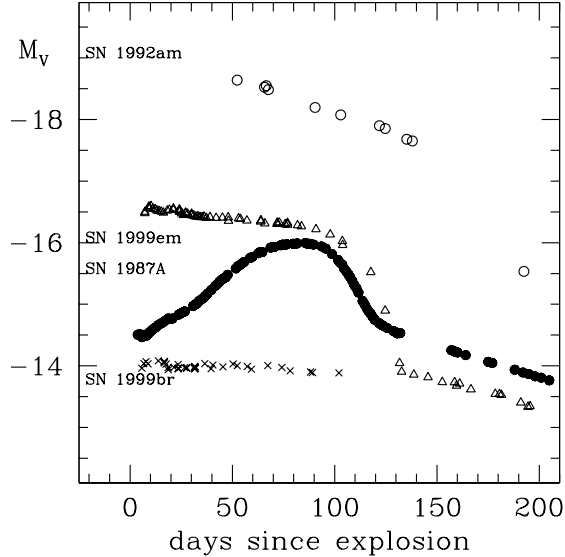}}
\caption{Absolute $V$ lightcurves of classical Type II SN.
The observations are from \cite{hamuy90} for SN~1987A;
from \cite{hamuy01a} for SN~1992am and SN~1999br; and
from \cite{suntzeff03} for SN~1999em.}
\label {LC1.fig}
\end{figure}

A sample of absolute $V$ lightcurves of classical  SNe~II is shown in Fig. \ref{LC1.fig}.
SN~1999em is among the best observed objects of this class which is characterized by
a long plateau ($\sim$110 days) of nearly constant luminosity. The plateau is followed
by a sudden drop in luminosity and a linear tail with a slope of $\sim$0.01 mag day$^{-1}$.
The plateau corresponds to the optically thick phase in which the SN has a well-defined
photosphere. Since the opacity is dominated by e$^-$ scattering, the photosphere lies
at the shell where H recombines. As the ejecta expands a recombination wave recedes 
through the envelope.  The end of the plateau corresponds to the time when the
photosphere reaches the He-rich envelope. Owing to much lower opacities 
the photosphere recedes faster and the luminosity drops promptly until the SN
becomes transparent. Up to this point the lightcurve is powered primarily by the internal
energy of the SN previously deposited by the shock wave that ensued from core collapse.
During the nebular phase, on the other hand, the lightcurve is powered
by the radioactive decay of $^{56}$Co $\rightarrow$ $^{56}$Fe,
at a rate corresponding to the e-folding time of the $^{56}$Co decay (111.26 days).
$^{56}$Co is the daughter of $^{56}$Ni (with a half-life of 6.1 days),
so the luminosity of the tail is determined by the amount of $^{56}$Ni
freshly synthesized in the explosion. Plateau SNe (SNe~II-P) such as SN~1999em comprise
the vast majority of spectroscopically classical SNe~II. This photometric class
was first identified by \cite{barbon79}. Two more examples are shown
in Fig. \ref{LC1.fig}: SN~1999br and SN~1992am. It is evident that there
is a great range ($\sim$5 mag) in luminosity within this group.

\begin{figure}[ht]
\centerline{\includegraphics[width=3.3in,height=3.3in]{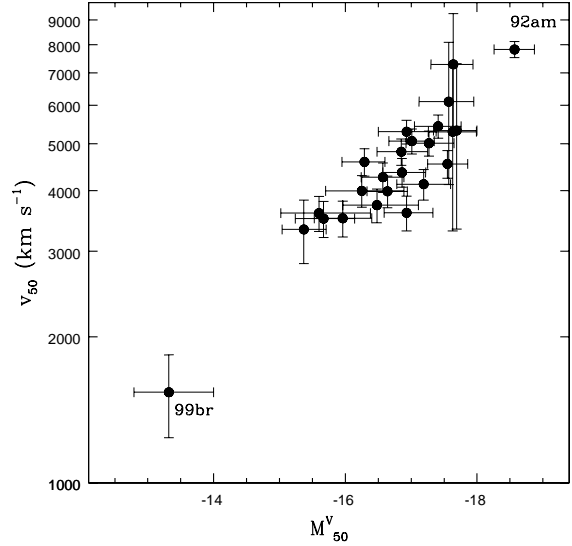}}
\caption{Expansion velocities from Fe II $\lambda$5169 versus
absolute $V$ magnitude, both measured in the middle of the plateau
(day 50) of 24 Type II SNe~II-P.}
\label{L_v.fig}
\end{figure}

SN~1987A is the most well-observed SN to date and its lightcurve (Fig. \ref{LC1.fig})
proves clearly different than that of SNe~II-P. It is characterized
by a steady rise during three months. After maximum light the SN displayed a
fast decline phase of $\sim$20 days, followed by a linear tail at the rate
expected for $^{56}$Co $\rightarrow$ $^{56}$Fe. Its peculiar shape has been
attributed to the relatively small radius of its blue supergiant
progenitor Sk 202-69 (\cite{Woosley87}). Unlike SNe~II-P which explode as red
supergiants (\cite{Arnett96}), most of the shock deposited energy in SN~1987A went
into adiabatic expansion, thus leading to a dimmer plateau and to
a lightcurve promptly powered by $^{56}$Ni $\rightarrow$ $^{56}$Co $\rightarrow$ $^{56}$Fe (\cite{Blinnikov00}).
The lightcurve shape of SN~1987A reflects the combination of an
ever decreasing deposition rate with an ever increasing escape
probability for the photons from the SN interior as the ejecta gets thinner.
SN~1998A (\cite{Woodings98}) and SN~2000cb (\cite{Hamuy01a}) are two other
clear examples of events with SN~1987A-like lightcurves and, hence, with
blue supergiant progenitors. The relatively compact progenitors of these SNe have
been attributed to low metallicities and to mass loss to a binary
companion prior to explosion.

Figures \ref{spec1.fig} and \ref{LC1.fig} suggest that SNe
with brighter plateaus have higher ejecta velocities, and viceversa.
Figure \ref{L_v.fig} shows expansion velocities versus plateau
luminosities for the 24 SNe~II-P having sufficient photometric and
spectroscopic data (see \cite{Hamuy03} for details). Despite the great diversity
displayed by SNe~II-P, these objects show a tight luminosity-velocity correlation.
This result suggests that, while the explosion energy
increases so do the kinetic and internal energies. In fact,
as shown in Fig. 2 of \cite{hamuy03}, the luminosity-velocity correlation is
also present in the theoretical models of \cite{litvinova85} (LN85).

\begin{figure}[ht]
\centerline{\includegraphics[width=3.3in,height=3.3in]{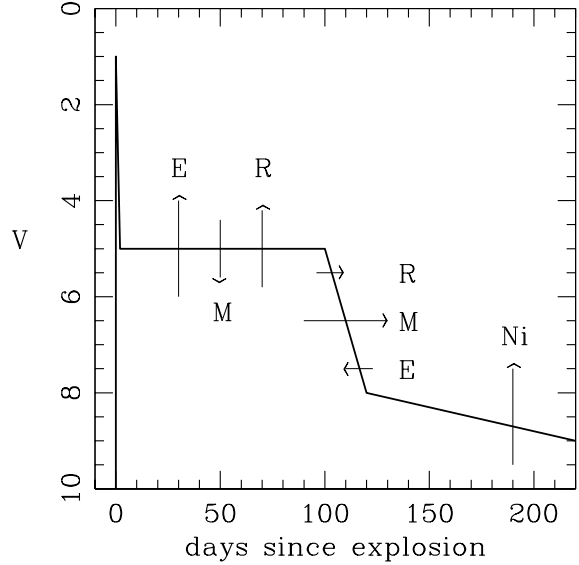}}
\caption{Effect of explosion energy ($E$), ejected  mass ($M$),
initial radius ($R$), and nickel mass ($Ni$) on the lightcurve
of a SN~II-P.}
\label{PLC.fig}
\end{figure}

\begin{figure}[ht]
\centerline{\includegraphics[width=4.0in,height=4.0in]{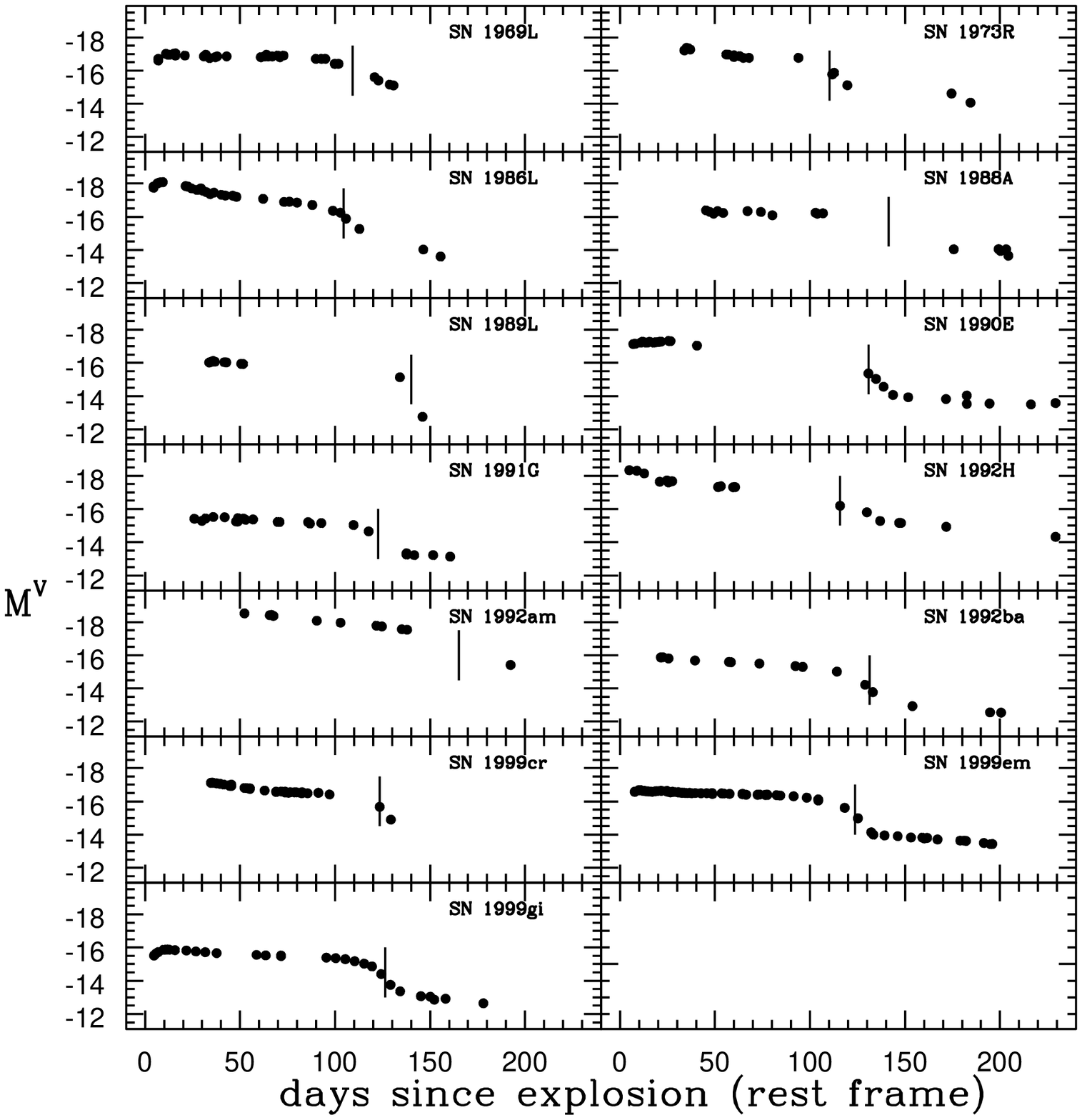}}
\caption{Extinction corrected absolute $V$-band lightcurves of
13 plateau SNe~II. The vertical bars indicate the end of the
plateau phase for each SN. The observations are from \cite{ciatti71}
for SN~1969L; \cite{ciatti77} for SN~1973R; \cite{phillips03} for SN~1986L;
\cite{ruiz90}, \cite{benetti91}, \cite{turatto93b} for SN~1988A;
\cite{schmidt03} for SN~1989L; \cite{schmidt93} and \cite{benetti94} for SN~1990E;
\cite{blanton95} for SN~1991G; \cite{tsvetkov94} and \cite{clocchiatti96a} for SN~1992H;
\cite{hamuy01a} for SN~1992am, SN~1992ba, and SN~1999cr; \cite{suntzeff03} and \cite{leonard02a}
for SN~1999em; and \cite{leonard02b} for SN~1999gi.}
\label{MLC.fig}
\end{figure}

Using the hydrodynamic models of LN85 it is possible to derive physical
parameters such as energy ($E$), ejected mass ($M$), and initial Radius ($R$)
for SNe~II-P. In such models the lightcurve is shaped by these parameters and 
their effect on the plateau phase is illustrated in Fig. \ref{PLC.fig}.
While the plateau luminosity is particularly sensitive to $E$, its
duration depends largely on $M$. The models of LN85 yield specific
calibrations for three observables: the plateau luminosity, its duration,
and the photospheric velocity, which can be used to solve for $E$, $M$, and $R$.
One problem with the LN85 models is that they do not cover a wide range
in energy and mass so the results derived from this calibration
often involve extrapolating their formulas. Although it is
necessary to expand the parameter space explored by LN85 before
we can firmly believe this method, in its current form
it can still provide useful insights on the nature of these objects.
As mentioned above, the luminosity of the radioactive tail can be used to find
the mass of $^{56}$Ni ($M_{Ni}$) synthesized in the explosion, assuming that all the $\gamma$-rays
from $^{56}$Co $\rightarrow$ $^{56}$Fe are fully thermalized in the interior.
Fig. \ref{MLC.fig} shows the 13 SNe~II-P for which there is sufficient
data to perform such analysis.

\begin{table}[ht]
\caption[] {Physical Parameters for Classical Type II Supernovae.}
\begin{tabular*}{\textwidth}{@{\extracolsep{\fill}}lcccc}
\sphline
\it SN & \it $Energy$                   & \it $Ejected$ $Mass$ & \it $Initial$ $Radius$   & \it $Nickel$ $Mass$   \cr
\it    & \it {($\times$10$^{51}$ ergs)} & \it {$(M_\odot)$}    & \it {$(R_\odot)$}        & \it {$(M_\odot)$}     \cr
\sphline
1969L      &  2.3$_{\rm -0.6}^{+0.7}$  & 28$_{\rm  -8}^{+11}$ & 204$_{\rm  -88}^{+150}$ &  0.082$_{\rm -0.026}^{+0.034}$ \cr
1973R      &  2.7$_{\rm -0.9}^{+1.2}$  & 31$_{\rm -12}^{+16}$ & 197$_{\rm  -78}^{+128}$ &  0.084$_{\rm -0.030}^{+0.044}$ \cr 
1986L      &  1.3$_{\rm -0.3}^{+0.5}$  & 17$_{\rm  -5}^{ +7}$ & 417$_{\rm -193}^{+304}$ &  0.034$_{\rm -0.011}^{+0.018}$ \cr
1988A      &  2.2$_{\rm -1.2}^{+1.7}$  & 50$_{\rm -30}^{+46}$ & 138$_{\rm  -42}^{ +80}$ &  0.062$_{\rm -0.020}^{+0.029}$ \cr
1989L      &  1.2$_{\rm -0.5}^{+0.6}$  & 41$_{\rm -15}^{+22}$ & 136$_{\rm  -65}^{+118}$ &  0.015$_{\rm -0.005}^{+0.008}$ \cr
1990E      &  3.4$_{\rm -1.0}^{+1.3}$  & 48$_{\rm -15}^{+22}$ & 162$_{\rm  -78}^{+148}$ &  0.062$_{\rm -0.022}^{+0.031}$ \cr
1991G      &  1.3$_{\rm -0.6}^{+0.9}$  & 41$_{\rm -16}^{+19}$ &  70$_{\rm  -31}^{ +73}$ &  0.022$_{\rm -0.006}^{+0.008}$ \cr
1992H      &  3.1$_{\rm -1.0}^{+1.3}$  & 32$_{\rm -11}^{+16}$ & 261$_{\rm -103}^{+177}$ &  0.129$_{\rm -0.037}^{+0.053}$ \cr
1992am     &  5.5$_{\rm -2.1}^{+3.0}$  & 56$_{\rm -24}^{+40}$ & 586$_{\rm -212}^{+341}$ &  0.256$_{\rm -0.070}^{+0.099}$ \cr
1992ba     &  1.3$_{\rm -0.4}^{+0.5}$  & 42$_{\rm -13}^{+17}$ &  96$_{\rm  -45}^{+100}$ &  0.019$_{\rm -0.007}^{+0.009}$ \cr
1999cr     &  1.9$_{\rm -0.6}^{+0.8}$  & 32$_{\rm -12}^{+14}$ & 224$_{\rm  -81}^{+136}$ &  0.090$_{\rm -0.027}^{+0.034}$ \cr
1999em     &  1.2$_{\rm -0.3}^{+0.6}$  & 27$_{\rm  -8}^{+14}$ & 249$_{\rm -150}^{+243}$ &  0.042$_{\rm -0.019}^{+0.027}$ \cr
1999gi     &  1.5$_{\rm -0.5}^{+0.7}$  & 43$_{\rm -14}^{+24}$ &  81$_{\rm  -51}^{+110}$ &  0.018$_{\rm -0.009}^{+0.013}$ \cr
\sphline
       &  & \it $SNe$ $from$ $Other$ $Sources$ & &  \cr
\sphline
1987A$^a$  &  1.7                      & 15                   &  42.8                   &  0.075 \cr
1997D$^b$  &  0.9                      & 17                   &  128.6                  &  0.006 \cr
1999br$^b$ &  0.6                      & 14                   &  114.3                  &  0.0016$_{\rm -0.0008}^{+0.0011}$ \cr
\sphline
\end{tabular*}
\begin{tablenotes}
$^a$ From \cite{arnett96}.

$^b$ From \cite{zampieri03}.
\end{tablenotes}
\label{PMII.tab}
\end{table}

Table \ref{PMII.tab} summarizes the resulting parameters and those independently
derived for SN~1987A (\cite{Arnett96}), SN~1997D and SN~1999br (\cite{Zampieri03}),
which reveals the following,

\noindent $\bullet$ There is a wide range in explosion energies, 
from 0.6 to 5.5 foes (1 foe = 10$^{51}$ ergs) among classical SNe~II.

\noindent $\bullet$ The ejected masses encompass a broad range between
14 and 56 $M_\odot$. Note that, while stars born with more than 8 $M_\odot$ 
can in principle undergo core collapse, they do not show up as classical SNe~II.
Perhaps they become white dwarfs (\cite{Heger03}) or they undergo significant
mass loss before explosion and are observed as other SN spectroscopic types (\cite{Chugai97a}).
Note also that stars as massive as 50 $M_\odot$, which
are expected to have strong stellar winds (\cite{Heger03}), seem able to retain
a significant fraction of their H envelope and explode as SNe~II,
perhaps owing to lower metallicities. Although these results prove interesting,
it must be mentioned that the derived masses are quite uncertain because
the LN85 calibration only extends up to 16 $M_\odot$. Of some concern
is the sharp contrast found between the ejected masses derived from the LN85
calibration for SN~1999gi and SN~1999em
and the values independently obtained by Smartt et al. (2001, 2002).
Based on upper limits of the luminosities of the progenitors of these two
nearby SNe from pre-discovery images and stellar evolutionary tracks,
they derived upper mass limits of 9$_{\rm -2}^{+3}$ and 12$_{\rm -1}^{+1}$ $M_\odot$
for SN~1999gi and SN~1999em, which prove at odds with the values of
43$_{\rm -14}^{+24}$ and 27$_{\rm  -8}^{+14}$ $M_\odot$ obtained
for these objects from the LN85 calibration. Part of the discrepancy
might arise from the LN85 calibration, but it could be due also to 
the distances adopted by Smartt et al. for the SN host galaxies
(in fact with the new Cepheid distance to SN~1999em the upper mass limit
rises to $\sim$20$M_\odot$; \cite{Leonard03}), or
uncertain stellar evolutionary models for massive stars.

\noindent $\bullet$ Except for SN~1987A, within the uncertainties
the initial radii correspond to those measured for red supergiants (\cite{Vanbelle99}),
which lends support to the view that the progenitors of SNe~II-P have extended
atmospheres at the time of explosion (\cite{Arnett96}).

\noindent $\bullet$ The Ni masses produced by these SNe vary by
a factor of $\sim$100, from 0.0016 to 0.26 $M_\odot$.

\begin{figure}[ht]
\centerline{\includegraphics[width=3.3in,height=3.3in]{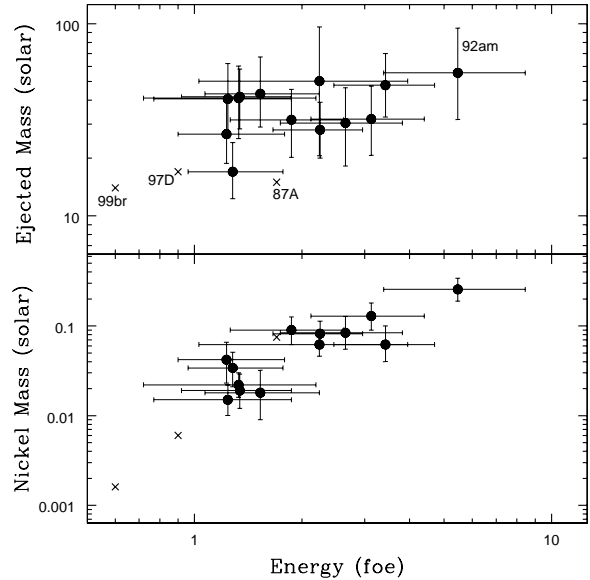}}
\caption{Ejected mass and nickel mass of classical SNe~II, as a function
of explosion energy. Filled circles correspond to the 13 SNe~II-P for which
the LN85 models could be applied, and the three crosses correspond to SN~1987A,
SN~1997D, and SN~1999br, which have been modeled in detail by \cite{arnett96}
and \cite{zampieri03}. }
\label{ME.fig}
\end{figure}

Fig. \ref{ME.fig} shows $M$ and $M_{Ni}$ as a function of $E$ for these 16 SNe~II.
Despite the large error bars, this figure reveals a couple of correlations.
First (top panel), the explosion energy appears to be correlated with the ejected mass, in the sense
that more energetic SNe eject greater masses. This suggests that the outcome
of the core collapse is somehow determined by the mass of the envelope, or that the mass
of the envelope is correlated with some property of the core (e.g. mass), with the latter
determining the outcome. Second (bottom panel), SNe with greater energies produce more nickel,
a result previously suggested by \cite{blanton95}.
This could mean that greater temperatures and more nuclear burning are reached in such SNe,
and/or that less mass falls back onto the neutron star/black hole in more energetic explosions.

\subsection{ Type IIdw Supernovae }

A distinct class of SNe~II can been identified which, unlike
classical SNe~II, are believed to be strongly interacting with a ``dense
wind'' produced by the SN progenitor prior to explosion.
These SNe have strong radio emission
caused by the interaction with the CSM (\cite{Chevalier98}). Models of the radio observations
imply high mass-loss rates $\sim$10$^{-4}$ $M_\odot$ $yr^{-1}$ for their
progenitors (\cite{Weiler02}). SNe~IIdw comprise  $\sim$30\% of all CCSNe and
$\sim$40\% of all SNe~II.

\begin{figure}[ht]
\centerline{\includegraphics[width=3.3in,height=3.3in]{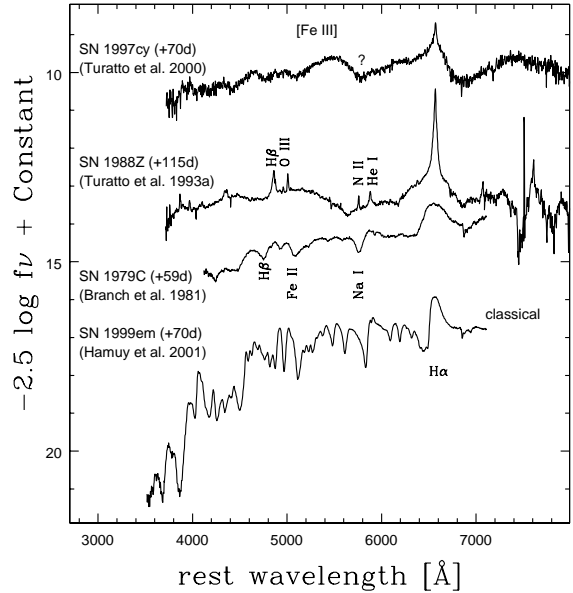}}
\caption{Optical spectra of SNe~IIdw obtained $\sim$3 months after explosion, compared to the classical Type II SN~1999em.}
\label{spec2.fig}
\end{figure}

Fig. \ref{spec2.fig} shows optical spectra for a sample of 
SNe~IIdw, compared to the classical SN~1999em. Evidently these SNe show
a strong degree of individuality, but they are unified by the
{\it lack of absorption} in the Balmer lines. Their spectra are
dominated by strong H$\alpha$ broad emission (SN~1979C), sometimes
with a superposed narrow (FWHM$\sim$200 km~s$^{-1}$) emission (SN~1988Z and SN~1997cy). 
One of most well-observed and recent additions to the IIdw events is SN 1998S (\cite{Leonard00}).
When the narrow component is present the SN is classified
as IIn (standing for ``narrow''; \cite{Schlegel90}). Occasionally a narrow P-Cygni
profile can be observed, such as in SN~1994aj (\cite{Benetti98}) and
SN~1996L (\cite{Benetti99}), in which case the SN is typed as IId
(the ``d'' stands for ``double'' profile).

The interaction of the SN envelope and the CSM is very difficult to model.
In the models of \cite{chugai97a} this collision produces an outer 
shock wave that propagates in the unshocked CSM, and an inner 
shock wave that propagates inward through the SN envelope. In between the two
shocks is located a cool and dense shell that produces broad emission lines
by excitation from the X-rays produced in the inner and outer shocks.
Broad emissions are also thought to arise in the undisturbed SN ejecta
excited by the X-rays. \cite{chugai97b} claims that the absence
of the absorption component of the H$\alpha$ profile observed in classical
SNe~II is a consequence of the excitation mechanism: while the atmospheres of the classical SNe~II are 
excited by internal energy of the explosion and by radioactivity,
SNe~IIdw are predominantly excited by the shocks. In these models
the narrow emission component in SNe~IIdw is thought to originate
in the undisturbed photoionized CSM. The narrow P-Cygni profile observed
in SNe~IId is attributed to the recombined unshocked CSM.

\begin{figure}[ht]
\centerline{\includegraphics[width=3.3in,height=3.3in]{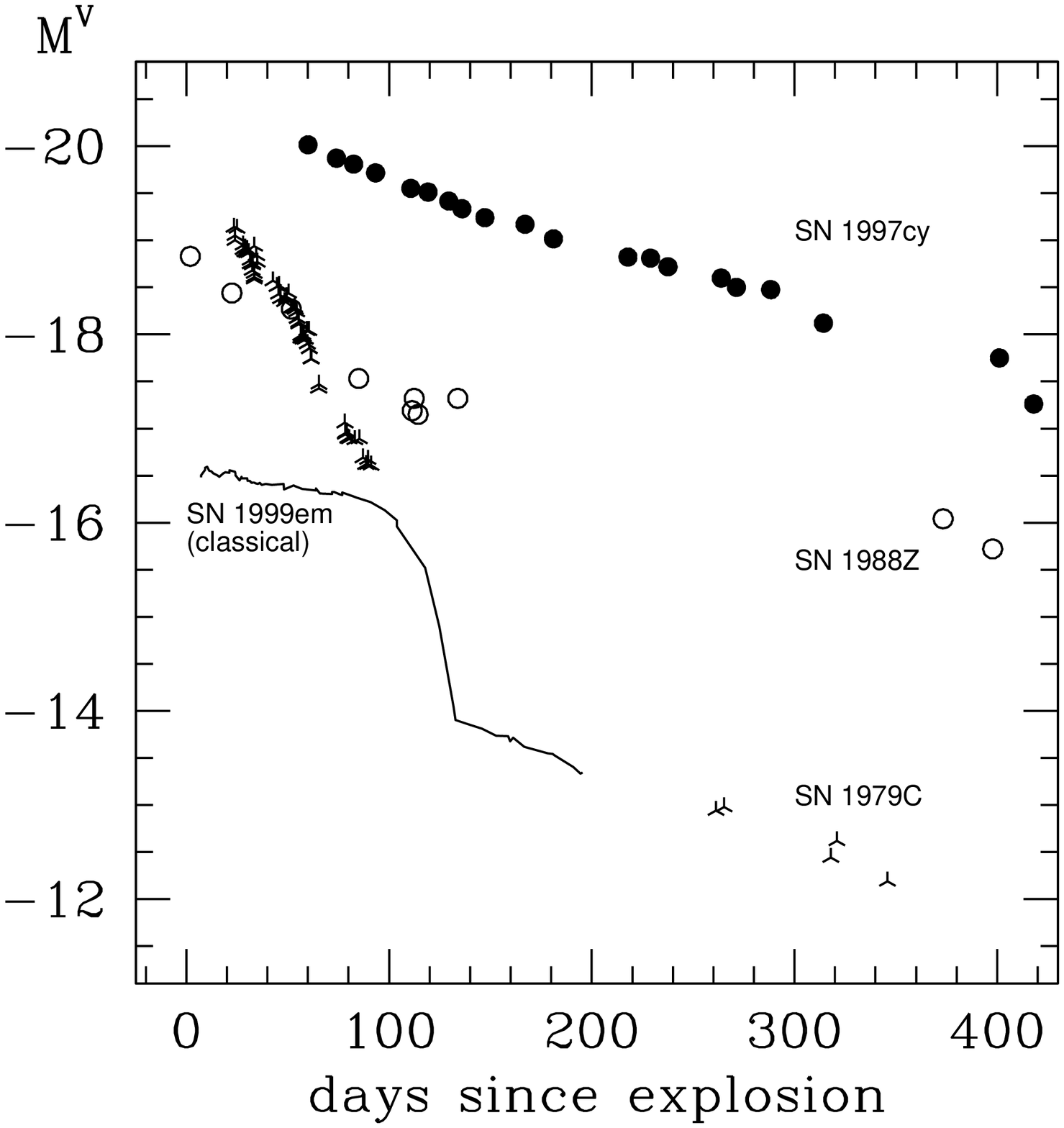}}
\caption{Absolute $V$ lightcurves of three SNe~IIdw, compared to the classical
Type II SN~1999em (solid line; \cite{Suntzeff03}). The observations are from
\cite{balinskaya80}, \cite{devaucouleurs81}, and \cite{barbon82b} for SN~1979C;
\cite{turatto93a} for SN~1988Z;
and \cite{germany00} for SN~1997cy.}
\label{LC2.fig}
\end{figure}

The lightcurves of some SNe~IIdw are compared to that of the classical
SN~1999em in Fig. \ref{LC2.fig}. The strong degree of individuality
seen in their spectra is also reflected in their lightcurves. SN~1979C
belongs to the photometric class of ``linear'' SNe introduced by \cite{barbon79}.
SN~1979C displayed a post-maximum decline phase at a rate of 0.04 mag day$^{-1}$
for $\sim$100 days, followed by a slower decline tail at $\sim$0.01 mag day$^{-1}$.
SN~1980K is another clear example of a linear event (\cite{Barbon82a}). SN~1988Z and SN~1997cy,
on the other hand, showed very different behaviors. They both faded slowly,
at 0.01 mag day$^{-1}$. This rate is not
very different than that of the classical plateau SN~1999em, but the main
difference is that this slow evolution in SN~1988Z and SN~1997cy extended for several
hundred days.

\begin{figure}[ht]
\centerline{\includegraphics[width=3.3in,height=3.3in]{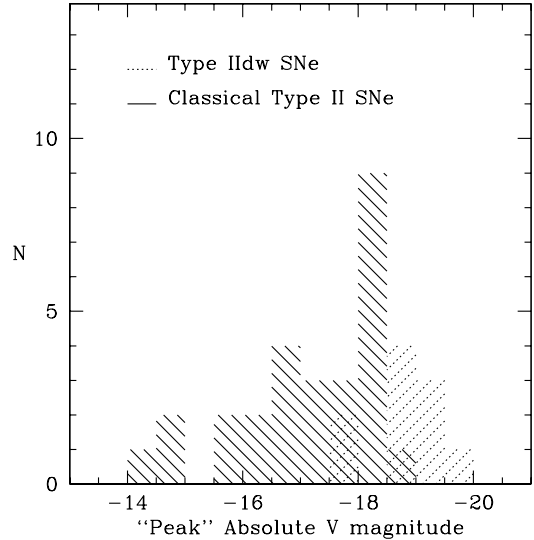}}
\caption{Distribution of absolute peak magnitudes of classical and
Type IIdw SNe.}
\label{absmag.fig}
\end{figure}

Despite the great photometric diversity among SNe~IIdw, these objects share the
property of being generally more luminous than the classical SNe~II.
This can be appreciated in Fig. \ref{absmag.fig}, which compares the distribution
of peak magnitudes of classical and Type IIdw SNe. The high luminosities of
Type IIdw SNe constitute another optical signature of the collision of
SN envelopes with dense winds, which efficiently converts kinetic energy into light.
Often, this mechanism predominates over the usual SN radiation mechanism,
thus leading to a powerful optical display (\cite{Chugai97a}).

\begin{table}[ht]
\caption[] {Physical Parameters for Type IIdw Supernovae.}
\begin{tabular*}{\textwidth}{@{\extracolsep{\fill}}lccccc}
\sphline
\it SN & \it $Energy$                   & \it $Ejected$        & \it $Initial$            & \it $Nickel$         & \it $Mass-loss$  \cr
\it    &                                & \it $Mass$           & \it $Radius$             & \it $Mass$           & \it $Rate$       \cr
\it    & \it {($\times$10$^{51}$ ergs)} & \it {$(M_\odot)$}    & \it {$(R_\odot)$}        & \it {$(M_\odot)$}    & \it {$(M_\odot~yr^{-1}$}) \cr
\sphline
1979C$^a$  &  1-2  & 6   & 6000  &  $<$0.1  & 1$\times$10$^{-4}$   \cr
1988Z$^b$  &  1    & $<$1&  ...  &  ...     & 1-7$\times$10$^{-4}$ \cr
1997cy$^c$ &  30   & 5   &  ...  &  ...     & 4.0$\times$10$^{-4}$ \cr
\sphline
\end{tabular*}
\begin{tablenotes}
$^a$ From \cite{blinnikov93} and \cite{weiler02}.

$^b$ From \cite{chugai94} and \cite{weiler02}.

$^c$ From \cite{turatto00} and \cite{nomoto00}.
\end{tablenotes}
\label{PMIIdw.tab}
\end{table}

Table \ref{PMIIdw.tab} summarizes physical parameters for three SNe~IIdw
derived from models involving circumstellar interaction. 
In general these SNe eject less mass than the classical SNe~II,
between 1-6 $M_\odot$. The models require large mass-loss rates before explosion.
Two of them exploded with normal energies compared to classical SNe~II,
but one case (SN~1997cy) was much more energetic than any
other SN~II.  \cite{chugai97a} proposes that 1) the progenitors of
SNe~IIdw are stars born with 8-10 $M_\odot$ that undergo significant
mass loss owing to helium shell flashes during the asymptotic red
supergiant branch stage, and 2) the large observed diversity among SNe~IIdw is
due to large variations in the wind density near the
SN progenitor. This hypothesis is consistent with the conclusions reached above
from the LN85 models, that the progenitors of the classical SNe~II are born with more than
14 $M_\odot$. 

\subsection{Type Ib and Ic Supernovae}

The distinguishing feature of Type Ib and Ic SNe is the lack
of conspicuous hydrogen spectral lines. Their progenitors are thought
to be massive stars that lose most of their H-rich (and perhaps
their He-rich) envelopes via strong winds or transfer to a binary
companion via Roche overflow.
Evidence for this hypothesis has been recently revealed by
the Type Ic SN~1999cq, which showed intermediate-width helium emission
lines caused by the interaction of the SN ejecta with an almost pure
helium wind lost by the progenitor (\cite{Matheson00}).
Type Ib and Ic SNe are as radio loud as SNe~IIdw, yet the 
mass-loss rates of their progenitors are lower, between 10$^{-5}$-10$^{-6}$ $M_\odot$ $yr^{-1}$
(\cite{Weiler02}). Approximately 25\% of all CCSNe fall in the SNe~Ib and SNe~Ic category.

\begin{figure}[ht]
\centerline{\includegraphics[width=3.3in,height=3.3in]{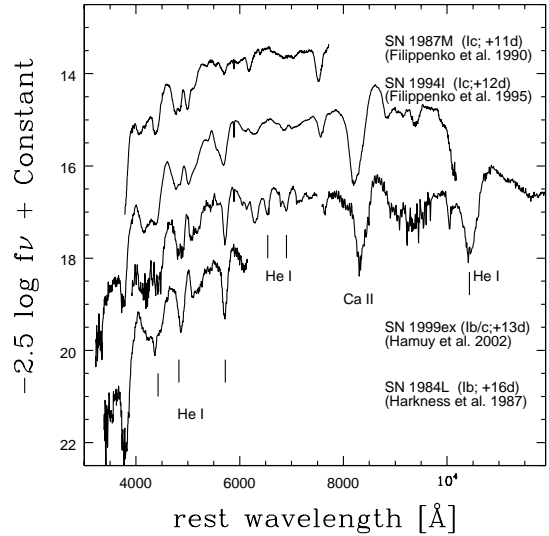}}
\caption{Optical spectra of Type Ib and Type Ic SNe, taken $\sim$14 days after maximum-light.}
\label{spec3.fig}
\end{figure}

Fig. \ref{spec3.fig} shows near-maximum optical spectra of various members of the Type I
core collapse family. SN~1984L is one of the prototypes of the Type Ib class,
which is characterized by strong He I $\lambda$$\lambda$4471, 5876, 6678, and 7065 lines.
These objects are not seen very often and comprise only $\sim$1\% of all CCSNe.
SN~1987M was the first well established member of the Ic family and SN~1994I
proves to be one of the best observed objects of this class. The strongest lines
are the Ca II H\&K and O I $\lambda$7774 absorptions, and the P-Cygni
profile of the Ca II IR triplet.

Significant effort has been put over recent years into determining the presence of helium
in the spectra of SNe~Ic in order to better understand the nature of these objects.
Recently \cite{matheson01} compiled and analyzed a large collection of spectra
of SNe~Ib and Ic obtained at Lick Observatory.
This study showed no compelling evidence for He in absorption in the spectra of SNe~Ic
and no gradual transition from the Ib to the Ic class, which supported the idea
that these objects are produced by different progenitors. By contrast, a closer look
at the SN~1994I spectrum suggests deeper troughs at the wavelengths of the He I lines, especially
at 4471, 4921, and 5876 Ang (\cite{Filippenko95,Clocchiatti96b}). The spectrum of SN~1999ex
shown in Fig. \ref{spec3.fig} reveals unambiguous evidence for He I absorptions of moderate
strength in the optical region, thus suggesting the existence of an intermediate Ib/c
case and a link between the Ib and Ic classes. 

The presence of hydrogen in SNe~Ib has been recently analyzed by \cite{branch02a}
by comparison to synthetic spectra. Their conclusion is that the 6300 Ang absorption feature
seen in the Type Ib SN~1954A, SN~1999di and SN~2000H is H$\alpha$, and that hydrogen
appears to be present in SNe Ib in general. A similar analysis led \cite{branch02b} to
conclude that H$\alpha$ is also present in the spectrum of the intermediate Type Ib/c SN~1999ex.
The presence of hydrogen in Type Ic SNe, on the other hand, is still questionable. 

The spectral models of \cite{branch02a} show that small optical depths
are required to fit the H and He lines and that a mild reduction in the optical
depths would make SNe~Ib look like SNe Ic.  This supports the concept that
SNe~Ib and SNe~Ic are not fundamentally different. Their spectroscopic
appearance may not even reflect the physical presence of H or He in their
atmospheres but only the way these elements are mixed in the envelopes.

\begin{figure}[ht]
\centerline{\includegraphics[width=3.3in,height=3.3in]{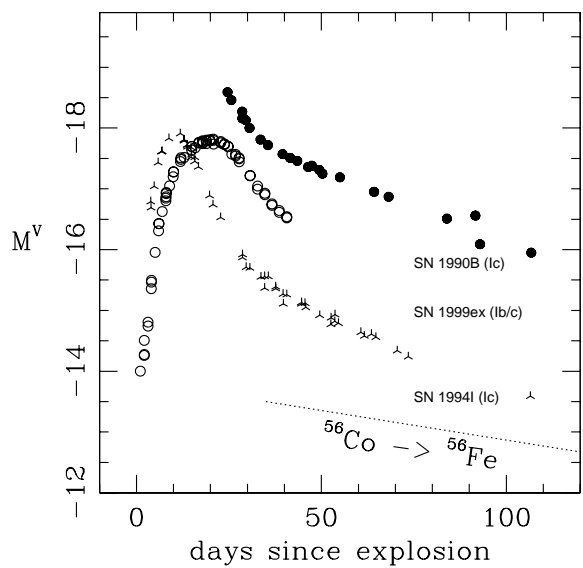}}
\caption{Absolute $V$ lightcurves of three Type I CCSNe. 
The observations are from
\cite{clocchiatti01} for SN~1990B;
\cite{richmond96} for SN~1994I;
and \cite{stritzinger02} for SN~1999ex.
The dotted line shows the lightcurve expected from $^{56}$Co $\rightarrow$ $^{56}$Fe.}
\label{LC3.fig}
\end{figure}

The lightcurves of some of the best observed Type I CCSNe are shown in Fig. \ref{LC3.fig}.
Evidently the lightcurves do not follow a single template, yet they share the
same common features: a rise during 2-3 weeks, a fast dimming during $\sim$30 days,
followed by a slow decline phase at a rate significantly greater than that
expected for the radioactive decay of $^{56}$Co $\rightarrow$ $^{56}$Fe (shown with
a dotted line).

Unlike classical SNe~II-P, the early-time lightcurves of SNe~I are
not powered by shock-deposited energy. This is attributed to the fact that
these objects have much more compact progenitors, so that
the energy deposited in the envelope by the central collapse is largely spent in adiabatic expansion,
which leads to a dimmer and brief plateau (\cite{Woosley87,Shigeyama90}).
Hence, from the very beginning their lightcurves are powered by radioactive heating.
While the peak is determined by the amount of $^{56}$Ni synthesized in the explosion,
the width depends on the ability of the photons to diffuse out from the SN interior.
The width of the lightcurve is determined by the diffusion time which 
increases with the envelope mass and decreases with expansion velocity.
The early-time lightcurve, therefore, provides useful constraints on the $^{56}$Ni mass,
envelope mass, and kinetic energy (\cite{Arnett96}). 
The great range in peak luminosities and lightcurve widths
displayed by SNe~I suggests a wide range in mass, velocity, and $^{56}$Ni.
The late-time decline rate reveals that a fraction of the $\gamma$-rays from
$^{56}$Co $\rightarrow$ $^{56}$Fe escape from the SN ejecta without being thermalized and,
therefore, can be used to quantify the degree of $^{56}$Ni mixing in the SN interior.
In the next section I summarize the physical parameters for these objects.

\subsection{Type Ic Hypernovae}

In the past five years three SNe (SN~1997ef, SN~1998bw, and SN~2002ap) have been
found to display very peculiar spectra compared to the standard types described above. Since
they show an overall similarity to each other, they seem to define
a new spectroscopic class of SNe.

\begin{figure}[ht]
\centerline{\includegraphics[width=3.3in,height=3.3in]{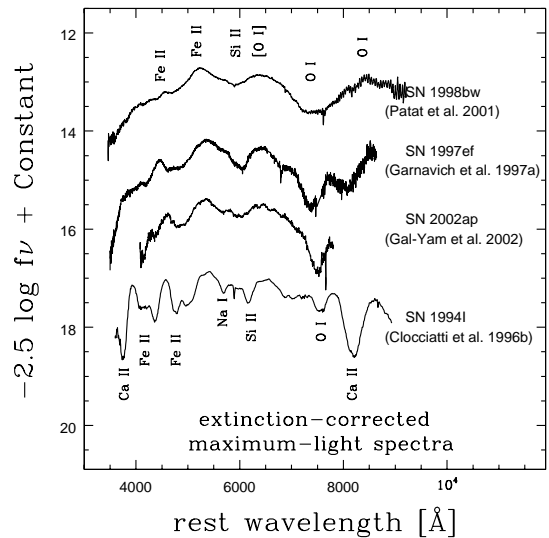}}
\caption{Maximum-light optical spectra of the three Type Ic hypernovae, compared to the normal Type Ic SN~1994I. }
\label{spec4.fig}
\end{figure}

Their maximum-light spectra are shown in Figure \ref{spec4.fig} along with the normal Type Ic SN~1994I.
Their main characteristic is that they are extremely smooth and featureless. They do not
show obvious signatures of hydrogen or helium. The few spectral features
displayed by these SNe are very broad and hard to identify without spectral modeling. The comparison
with SN~1994I suggests that these weird SNe are Type Ic events but with
higher than normal expansion velocities. This would lead to significant
Doppler broadening and blending of the spectral lines and, hence, to
featureless spectra. Judging from the line widths the velocities 
increase from SN~2002ap, to SN~1997ef, and SN~1998bw.
Their unusual expansion velocities suggest that
these objects are hyper-energetic so they are collectively called
``hypernovae''. Some people prefer to call them Type Id events.

\begin{figure}[ht]
\centerline{\includegraphics[width=3.3in,height=3.3in]{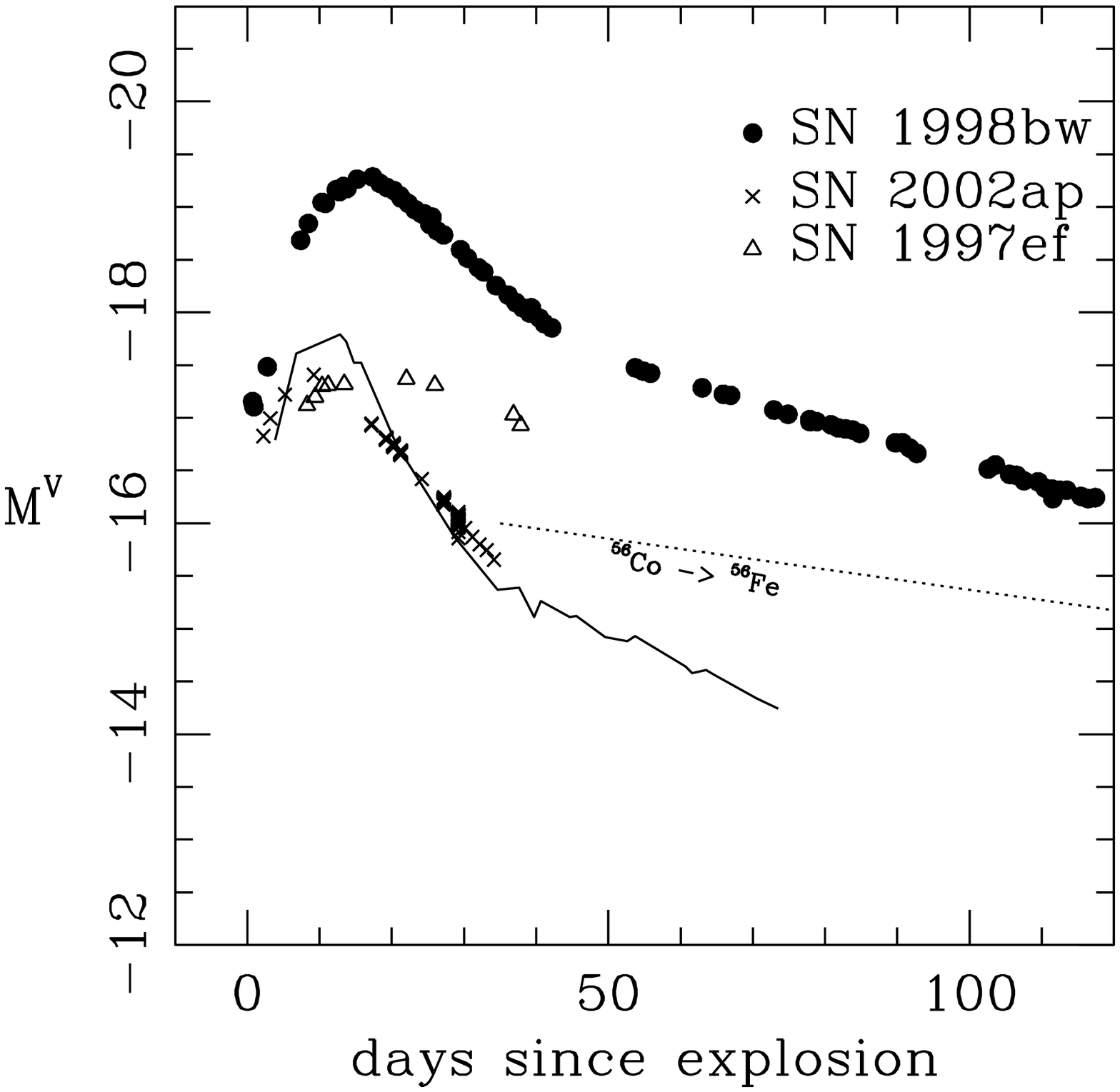}}
\caption{Absolute $V$ lightcurves of the three Type Ic hypernovae, compared to
the normal Type Ic SN~1994I (solid line). The observations are from
\cite{garnavich97a} and \cite{garnavich97b} for SN~1997ef;
\cite{galama98}, \cite{mckenzie99}, \cite{sollerman00}, and \cite{patat01} for SN~1998bw;
and \cite{galyam02} for SN~2002ap.}
\label{LC4.fig}
\end{figure}

The absolute $V$ lightcurves of the three hypernovae are shown in Fig. \ref{LC4.fig}
along with the normal Type Ic SN~1994I (solid line). They all share the generic shape of
normal SNe~Ib and SNe~Ic and, like these objects, the hypernovae show
great dispersion in lightcurve widths and luminosities. 
While both SN~1997ef and SN~2002ap showed normal luminosities, the
former had a broad lightcurve and the latter evolved much faster.
SN~1998bw showed a slow luminosity evolution and was clearly overluminous
compared to the other Type Ic events, which suggests a large $^{56}$Ni production.

SN~1998bw was not only remarkable for its great expansion velocities and luminosity,
but also because it exploded at nearly the same location and time as GRB980425 (\cite{Galama98}).
No GRBs were detected at the position of the other two hypernovae, on the other hand.
Since GRBs could arise in relativistic jets (\cite{Macfadyen01}), it could well be
that all hypernovae produce GRBs and that only observers
within a range of viewing angles from the jet axis can detect them.

\begin{table}[ht]
\caption[] {Physical Parameters for Type Ib and Type Ic Supernovae.}
\begin{tabular*}{\textwidth}{@{\extracolsep{\fill}}lcccccc}
\sphline
\it SN & \it $Type$ & \it $Energy$                   & \it $Ejected$        & \it $Initial$            & \it $Nickel$         & \it $Reference$  \cr
\it    &            &                                & \it $Mass$           & \it $Radius$             & \it $Mass$           &                  \cr
\it    &            & \it {($\times$10$^{51}$ ergs)} & \it {$(M_\odot)$}    & \it {$(R_\odot)$}        & \it {$(M_\odot)$}    &                  \cr
\sphline
1983I      & Ic  & 1.0  & 2.1 & 3.7   &  0.15  & \cite{Shigeyama90}   \cr
1983N      & Ib  & 1.0  & 2.7 & 3.0   &  0.15  & \cite{Shigeyama90}   \cr   
1984L      & Ib  & 1.0  & 4.4 & 1.9   &  0.15  & \cite{Shigeyama90}   \cr   
1994I      & Ic  & 1.0  & 0.9 & ...   &  0.07  & \cite{Nomoto00}      \cr
2002ap     & Ic  & 7.0  & 3.75& ...   &  0.07  & \cite{Mazzali02}     \cr
1997ef     & Ic  & 8.0  & 7.6 & ...   &  0.15  & \cite{Nomoto00}      \cr
1998bw     & Ic  &60.0  &10.0 & ...   &  0.50  & \cite{Nomoto00}      \cr   
\sphline
\end{tabular*}
\label{PMIbc.tab}
\end{table}

Nomoto and collaborators have modeled SNe~Ib as helium stars that lose
their hydrogen envelopes prior to explosion by mass transfer to a binary
companion, and SNe~Ic as C/O bare cores that lose their helium envelopes
in a second stage of mass transfer. In all cases they assume that these
SNe are spherically symmetric. Given the high levels of polarization
displayed by some of these objects (e.g. \cite{Wang01}, \cite{Leonard02c}), this might
not be an accurate approximation. Keeping in mind this caveat we list
in Table \ref{PMIbc.tab} their results and the physical parameters
they derive for the three hypernovae and four normal Type Ib and Ic events.

\begin{figure}[ht]
\centerline{\includegraphics[width=3.3in,height=3.3in]{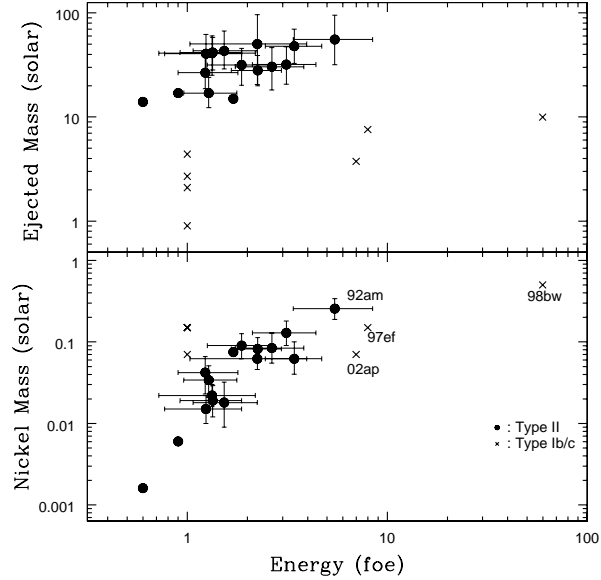}}
\caption{Ejected  mass and nickel yield of CCSNe, as a function of explosion energy.
Filled circles are the same 16 classical SNe~II shown in Fig. \ref{ME.fig}, 
and crosses correspond to the seven SNe~Ib and SNe~Ic listed in Table \ref{PMIbc.tab}.}
\label{ME_1.fig}
\end{figure}

Fig. \ref{ME_1.fig} shows how the ejected  masses and $^{56}$Ni yields
vary with explosion energy for the seven Type Ib and Type Ic SNe (crosses) listed in Table \ref{PMIbc.tab},
along with the 16 classical SNe~II (filled circles) shown in Fig. \ref{ME.fig}.
The top panel reveals that SNe~Ib and SNe~Ic appear to follow the same pattern shown by classical SNe~II,
namely, that more energetic SNe eject greater masses.
The main difference between both subtypes, of course, is the vertical offset caused by
the strong mass loss suffered by SNe~Ib and SNe~Ic prior to explosion. 
This suggests that the mass of the envelope is one of the driving parameters
of the explosion for CCSNe, or that it is correlated with some other
property of the core, with the latter determining the outcome of the collapse.

These plots permit one to appreciate that the distinguishing characteristic
of the three hypernovae is the great explosion energy, between 7 and 60 foes,
far above the normal Ib and Ic events which are characterized by energies $\sim$1 foe
(note, however, that this conclusion is based on spherically 
symmetric models; \cite{hoflich99} showed that the energy of SN~1998bw
could be as low as $\sim$2 foes when an ellipsoidal geometry for the ejecta is assumed).
When the whole sample of CCSNe is considered
there is a continous distribution of energies below 8 foes. 
Within this regime it appears that some Type II events such as SN~1992am
reach explosion energies comparable to that of the hypernovae
SN~1997ef and SN~2002ap. {\it This demonstrates that, in 
terms of energy, the definition of hypernova is not clear cut.}
Whether the energy distribution is continuous above
8 foes remains to be seen when more data become available. This will permit us to
understand if SN~1998bw belongs to a separate class of object or if it just lies
at the extreme of the family of CCSNe. At the moment it is fair to
say that there is only one firm SN/GRB association and, within
the SN context, this object was clearly exceptional regarding its explosion energy.

The bottom panel reveals that SN~1998bw was not only remarkable in
explosion energy (60 foes), but also in a high nickel yield (0.5 $M_\odot$).
Despite their greater than normal energies, none of the other two Type Ic hypernovae
(SN~1997ef and SN~2002ap) produced unusually higher nickel masses
compared to the other CCSNe. In fact the Type II SN~1992am ejected even
more $^{56}$Ni (0.26 $M_\odot$) than these two hypernovae. 
This suggests that there is continuum in $^{56}$Ni yields
below 0.3 $M_\odot$ for all CCSNe. More data are needed to ascertain
whether this continuity extends up to SN~1998bw.

Leaving aside SN~1998bw, it proves interesting that all other CCSNe
share the same location in the $^{56}$Ni/energy plane. Since both of these parameters carry
imprints from the physical conditions very close to the center
of the explosion, this suggests that the physics of the core and explosion
mechanisms of all CCSNe are not be fundamentally different.

\section{Summary and Discussion}

\noindent $\bullet$ The general properties of the different CCSNe mentioned
above are summarized in Table \ref{summary.tab}. 

\begin{table}[ht]
\caption[] {General Properties of Core Collapse Supernovae.}
\begin{tabular*}{\textwidth}{@{\extracolsep{\fill}}ccccc}
\sphline
\it $Spectroscopic$ & \it $Relative$      & \it $Photometric$ & \it $Mass-loss$                   & \it $Ejected$  \cr
\it $Type$          & \it $Frequency$     & \it $Type$        & \it $Rate$                        & \it $Mass$     \cr
\it                 & \it $(Observed)$$^a$& \it               & \it (10$^{-6}$ $M_\odot~yr^{-1})$ & \it $(M_\odot)$\cr
\sphline
II             & 45\%      & plateau,87A-like  & low                    & 14-56 \cr
IIdw           & 30\%      & linear,slow       & 10-700                 & $<$6  \cr
Ib+Ic          & 25\%      & bell-like         & 1-26                   & 1-10  \cr
\sphline
\end{tabular*}
\label{summary.tab}
\begin{tablenotes}
$^a$ Based on the LOTOSS (\cite{Li00}) discoveries reported to the IAU Circulars in 2001.
No attempt has been made here to correct the observed frequencies from selection biases.
The intrinsic SN rates must await the analysis of the observation history of the LOTOSS
galaxy sample.

\end{tablenotes}
\end{table}

\noindent $\bullet$ There are clear examples of intermediate spectroscopic cases in the CCSN family.
With intermediate He I line strengths, SN~1999ex provides a
clear link between the Ib and Ic classes (\cite{Hamuy02}). The metamorphosis of SN~1993J from a Type II SN
into a Ib event (\cite{Filippenko93}) provides evidence for an intermediate IIb case. 
This is strong evidence for a continous spectroscopic sequence (II-IIb-Ib-Ib/c-Ic) among CCSNe,
which reflects the ability of the SN progenitors to retain their H-rich and He-rich envelopes
prior to explosion. Another factor that defines the appearance of a SN is the density of the
medium in which they explode, which is determined by the history of mass-loss of the SN progenitor.

\noindent $\bullet$ There is a wide range in explosion energies (0.6-60 foes), ejected 
masses, and $^{56}$Ni yields (0.0016-0.5 $M_\odot$) among CCSNe, even within the same spectroscopic type.
Classical SNe~II show a trend in the sense that SNe with greater
envelope masses produce more energetic explosions (Fig. \ref{ME_1.fig}). Type Ib and Ic 
appear to follow the same pattern. This suggests that the mass of the
envelope is one of the driving parameters of the explosion, or that it is
correlated with some other property of the core, with the latter determining
the outcome of the collapse.

\noindent $\bullet$ Type II, Ib, and Ic SNe share
the same location in the $^{56}$Ni/energy plane. Since both of these parameters carry
imprints from the physical conditions very close to the center of the explosion,
this suggests that the physics of the core and explosion mechanisms of all CCSNe
are not be fundamentally different, regardless of the external appearance of these objects.

\noindent $\bullet$ At the moment it is fair to say that there is only one firm
SN/GRB association (SN~1998bw), and this object was clearly exceptional
regarding energy and nickel production within the SN context.

\begin{acknowledgments}

I am very grateful to Dave Branch, Inma Domínguez, Doug Leonard, and Schuyler van Dyk
for reading a draft of this review, which led to useful discussions
and a significant improvement of the paper.
Support for this work was provided by NASA through Hubble Fellowship grant HST-HF-01139.01-A
awarded by the Space Telescope Science Institute, which is operated by the Association
of Universities for Research in Astronomy, Inc., for NASA, under contract NAS 5-26555.
\end{acknowledgments}

\end{document}